\documentclass[prl,showpacs,twocolumn,superscriptaddress,floatfix,amsmath,raggedbottom]{revtex4}
\usepackage{amsmath}
\usepackage{amsthm}
\usepackage{amssymb}
\usepackage{amscd}
\usepackage{eucal}
\usepackage{bm}
\usepackage{graphicx}
\usepackage{subfigure}
\usepackage{color}
\usepackage[super]{nth}
\usepackage{dsfont}
\usepackage{hyperref}
\usepackage{stmaryrd}
\newcommand{\idop}{\mathds{1}}

\newcommand{\bb}[0]{\begin{eqnarray}}
\newcommand{\ee}[0]{\end{eqnarray}}

\def\dbar{\,{\mathchar'26\mkern-12mu d}}

\newcommand{\ket}[1]{| #1 \rangle}
\newcommand{\bra}[1]{\langle #1 |}

\begin{document}
\title{The role of quantum measurement in stochastic thermodynamics}
\author{Cyril Elouard}
\author{David A. Herrera-Mart\'i}
\affiliation{Institut N\'eel, UPR2940 CNRS and Universit\'e Grenoble Alpes, avenue des Martyrs, 38042 Grenoble, France.}
\author{Maxime Clusel}
\affiliation{Laboratoire Charles Coulomb, UMR5221 CNRS and Universit\'e de Montpellier, place E. Bataillon, 34095 Montpellier, France.}
\author{Alexia Auff\`eves}
\affiliation{Institut N\'eel, UPR2940 CNRS and Universit\'e Grenoble Alpes, avenue des Martyrs, 38042 Grenoble, France.}

\date{\today}
\begin{abstract}
This article sets up a new formalism to investigate stochastic thermodynamics in the quantum regime, where stochasticity and irreversibility primarily come from quantum measurement. In the absence of any bath, we define a purely quantum component to heat exchange, that corresponds to energy fluctuations caused by measurement back-action. Energetic and entropic signatures of measurement induced irreversibility are then investigated for canonical experiments of quantum optics, and the energetic cost of counter-acting decoherence is characterized on a simple state-stabilizing protocol. By placing quantum measurement in a central position, our formalism contributes to bridge a gap between experimental quantum optics and quantum thermodynamics.  \end{abstract}
\pacs{
05.30.-d, 05.40.-a, 05.70.Ln, 05.30.-d, 42.50.Lc
}
\maketitle

\begin{figure}[t]
\begin{center}
\includegraphics[scale=0.32]{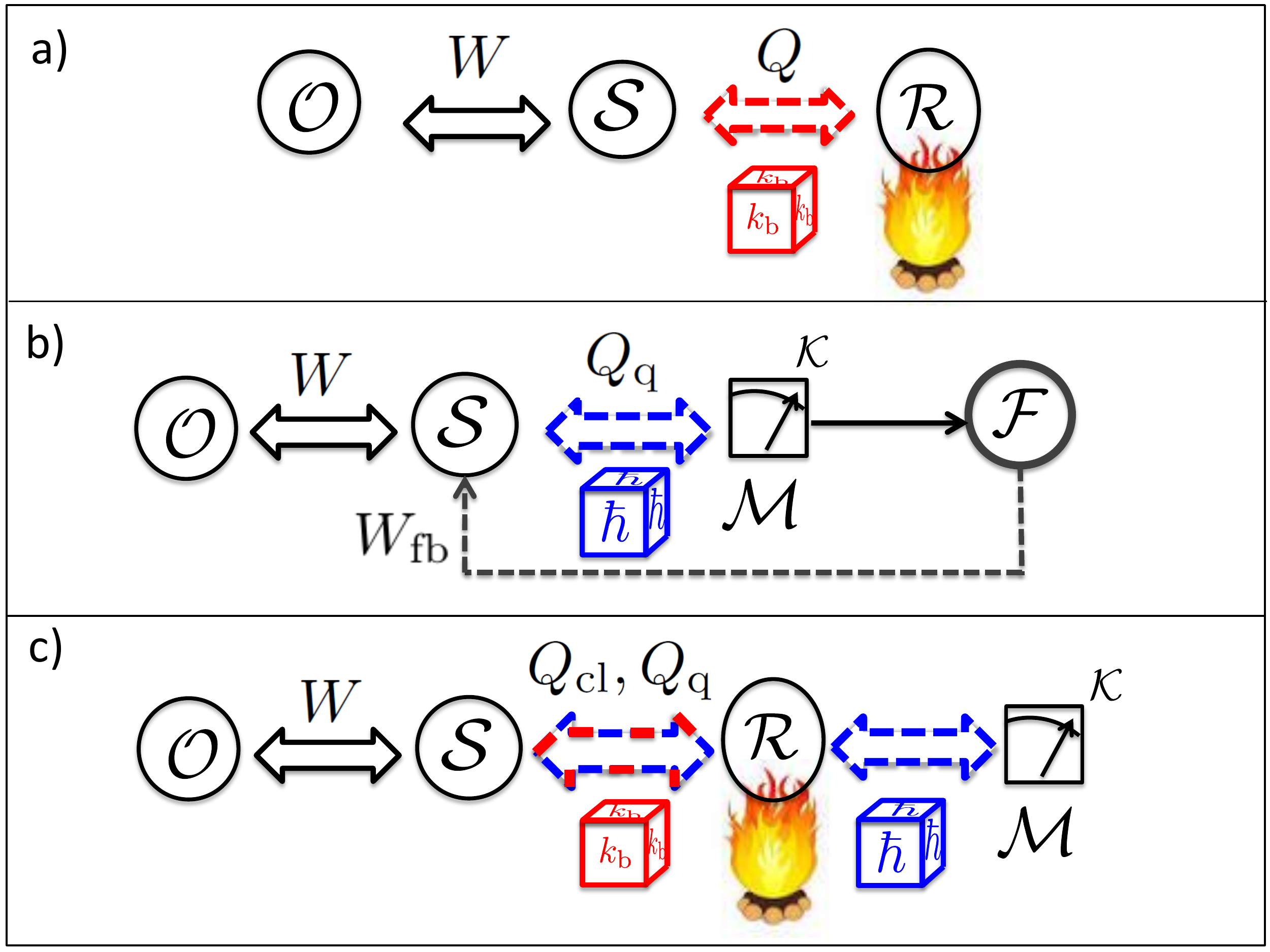}
\end{center}
\caption{a): The scenery of thermodynamics: A system ${\cal S}$ deterministically exchanges work $W$ with an operator/battery ${\cal O}$ and stochastically exchanges heat $Q$ with a bath/reservoir ${\cal R}$. The system's evolution is randomly perturbed by the thermal bath, which is symbolized by the dice k$_\mathrm{b}$ .  b): ``Thermodynamics without bath'': A driven quantum system undergoes projective quantum measurements at discrete times. Here the evolution of the system is randomized by the measuring device ${\cal M}$, which is symbolized by the dice $\hbar$. Energetic fluctuations induced by quantum measurement are identified with some heat exchange of quantum nature $Q_\text{q}$. In the case of a stabilizing protocol, this quantum heat exchange must be exactly compensated by the work $W_\text{fb}$ performed by a feedback source ${\cal F}$.
c): Quantum open systems: A driven system ${\cal S}$ is coupled to a Markovian reservoir ${\cal R}$, itself monitored by a measuring device ${\cal M}$. The exchanges of heat can be of classical ($Q_\text{cl}$) or of quantum nature ($Q_\text{q}$), and respectively correspond to the classical and quantum energy fluctuations induced by the coupling to the reservoir.}
\label{fig1}\end{figure}

Thermodynamics arose in the 19th century as a powerful theory to optimize thermal engines \cite{Carnot-24}, i.e. the extraction and storage of energy from thermal baths into batteries, by exploiting the transformations of a driven working fluid (Fig.\ref{fig1}a). This initially applied area of physics gave rise to fundamental concepts like thermodynamic time arrow: The irreversibility of a transformation is quantified by the produced entropy, which can only increase according to the second law of thermodynamics. In this classical framework, irreversibility and work extraction are tightly related, optimal work extraction being reached for reversible transformations.

Later on, stochastic thermodynamics extended these results at the microscopic level: heat, work and entropy are now defined for single trajectories followed by the system in its phase space \cite{Sekimoto-10, Seifert-08}. In particular, this framework accounts for the robustness of thermodynamic irreversibility, despite the reversibility of the physical laws at the microscopic scale  \cite{Lebowitz-93,Jarzynski-11}:  Because of the coupling to a stochastic entity like the thermal bath, the system's evolution is randomly perturbed, which breaks its reversibility. Time {reversal leads} to define the entropy produced in a single trajectory, and to the central Fluctuation Theorem (FT): The second law of thermodynamics and other celebrated FTs such as Jarzynski \cite{Jarzynski-97} and Crooks \cite{Crooks-99} relations then appear as particular cases of this central FT, highlighting the unifying strength of stochastic thermodynamics. 

Recently, these results have started being revisited in the quantum regime, where the working fluids, baths and batteries of thermodynamics are quantum entities.  This emerging framework brings out new questions, e.g. related to the work value of quantum coherence and entanglement  \cite{Del-Rio-11,Kammerlander-16, Feist-15,Lostaglio-15}, or the nature of irreversibility at the quantum scale \cite{Batalhao-15,Auffeves-15}. Extensions of the second law, especially of FTs, to the quantum regime, have been studied in various situations \cite{Esposito-09, Campisi-RMP-11}, while quantum trajectories were proposed as a promising operational tool to record such quantum FTs \cite{Horowitz-12,Horowitz-13,Hekking-13}. In this picture, quantum measurement most often plays an informational role, such as in Maxwell's demon setups \cite{Koski-15}. However quantum measurement is also a stochastic process that randomizes a quantum system's evolution, causing irreversibility \cite{Manzano-15}. While in classical thermodynamics, irreversibility has a clear energetic imprint, e.g. that alters the efficiency of heat engines, in quantum thermodynamics, energetic signatures related to measurement induced irreversibility and to the erasure of quantum coherences have remained elusive so far. 

Here we suggest a new framework for quantum thermodynamics, where energetic aspects of such genuinely quantum irreversibility become easily understandable
in terms of operational quantities. Our approach is based on the idea that a Markovian bath and a measuring apparatus in quantum mechanics play similar roles, i.e. are sources of stochasticity and irreversibility in the system's evolution. Our first goal is thus to build a ``thermodynamics without bath'', where stochasticity primarily comes from quantum measurement: This corresponds to the textbook situation of quantum mechanics where a driven system is solely coupled to a measuring device (Fig.\ref{fig1}b). We introduce new definitions for stochastic thermodynamic quantities, and especially introduce the key concept of quantum heat, which is identified with stochastic energy fluctuations taking place during a quantum measurement. Building on this analogy, we then extend our framework to the case of quantum open systems (Fig.\ref{fig1}c), where quantum thermodynamics was originally developed. Finally, we exploit this formalism to analyze energy, entropy and information transfers in two standard situations of quantum optics: The spontaneous emission of a Qubit and a state-stabilizing protocol. \\

\noindent \textbf{Results}\\
\noindent\textbf{Measurement based stochastic thermodynamics.}
%\noindent{\it Thermodynamics without bath.} 
We first introduce new definitions for thermodynamic quantities, considering the ideal situation pictured in Fig.\ref{fig1}b: A quantum system ${\cal S}$ is studied between the initial time $t_0$ and the final time $t_N$. ${\cal S}$  is driven by the Hamiltonian $H_\mathrm{s}(t)$, and undergoes projective measurements from the device ${\cal M}$ of eigenbasis $\{ \ket{m_{\cal K}} \}$ and eigenvalues $\{m_{\cal K}\}$. The set of eigenvalues is taken as discrete and their number is bounded by the dimension of the system's Hilbert space ${\cal N}_\text{s}$. The measurements are performed at discrete times $\{ t_n \}_{1\leq n \leq N}$, defining a set of stochastic records $\{ m_{{\cal K}_\gamma(t_n)} \}$. Finally, the measurement basis can change in time, but we shall not explicitly write this time-dependence. Let us suppose the initial state is a known pure state $\ket{\psi_0}$. If the measurement outcomes are read, the system remains in a pure state $\ket{\psi_\gamma(t)}$ at any time:  At time $t_n$, the system's state is stochastically projected on the pure state $\ket{\psi_n} = \ket{m_{{\cal K}_\gamma(t_n)}}$, and until time $t_{n+1}$, the system follows a Hamiltonian evolution $\ket{\psi_\gamma(t)} = {\cal G}(t, t_n)\ket{\psi_n}$. We have introduced ${\cal G}(v,u)= \mathfrak{T}\exp(-i\int_{u}^{v} dt H_\text{s}(t))$ as the evolution operator between time $u$ and $v$, $\mathfrak{T}$ being the time-ordering operator. $\ket{\psi_\gamma(t)}$ features an elementary quantum trajectory $\gamma$, that is perfectly defined by the measurement outcomes and the knowledge of the applied Hamiltonian. This trajectory is the quantum analog of the classical trajectories studied in stochastic thermodynamics, but in the present case the stochasticity primarily comes from quantum measurement. 

If the measurements are not read, the system's state is described by the density matrix $\rho_\mathrm{s}(t)$, which is recovered by averaging the system's stochastic states over all possible trajectories, at any time: $\rho_\mathrm{s}(t) = \langle \ket{\psi_\gamma(t)} \bra{\psi_\gamma(t)} \rangle_\gamma= \sum_\gamma P_\mathrm{d}[\gamma] \ket{\psi_\gamma(t)}\bra{\psi_\gamma(t)}$. We have introduced the probability of the trajectory $\gamma$: $P_\mathrm{d}[\gamma] = P_\mathrm{d}[\gamma|\psi_0] p_\mathrm{d}(\psi_0)$,  where $P_\mathrm{d}[\gamma|\psi_0] = \Pi_{n=0}^{N-1} |\langle \psi_{n+1}| {\cal G}(t_{n+1},t_{n})| \psi_n \rangle|^2$.
$p_\mathrm{d}(\psi_0)$ stands for the probability of the pure initial state, and equals $1$ if the state is known. It verifies  $p_\mathrm{d}(\psi_0)<1$ if the state is picked from a statistical mixture of orthogonal states $\{ \psi_k \}_{k=0}^{{\cal N}_\text{s}-1} = {\cal B}_0$: Such mixture can be prepared, e.g. by performing some projective measurement of an unknown state in the basis ${\cal B}_0$ before the transformation starts.  After a measurement has been performed at time $t_k$, the system's mean density matrix is diagonal in the eigenbasis of ${\cal M}$. In particular at $t=t_N$, it writes
 \bb \label{eq:dens}
 \rho_\text{s} (t_N) = \sum_{{\cal K}=1}^{{\cal N}_\text{s}}\pi_{\cal K} \ket{m_{\cal K}}\bra{m_{\cal K}}.
 \ee
\\

\label{p:MeasurementIrr}
\noindent \textbf{Measurement induced irreversibility.} The quantum trajectory picture highlights the irreversible character of a quantum measurement: Starting from the pure state $\ket{\psi_0}$ and applying the above protocol, we end up in the final state $ \ket{\psi_\gamma(t_N)} = \ket{\psi_N}$. Other trajectories can lead to the same final state, whose final probability equals $\pi_{{\cal K}_\gamma(t_N)}$.  Reciprocally, we consider the reverse protocol defined by picking the initial state $\ket{\psi_N}$ with probability $p_\text{r}(\psi_N)=\pi_{{\cal K}_\gamma(t_N)}$, applying the time-reversed Hamiltonian $H_\mathrm{s}(t_N-t)$ of corresponding evolution operator ${\cal G}^\text{r}$, and performing discrete measurements at times $t_{N-n}$: Such reverse protocol generally does not prepare back the state $\ket{\psi_0}$. The irreversibility associated to $\gamma$ is quantified by the entropy $\Delta_\mathrm{i}s[\gamma]$ produced along the trajectory: Denoting $P_\text{r}[\gamma_\text{r}]$ the probability of the time-reversed trajectory {$\gamma_\text{r} : \ket{\psi_\gamma(t_N-t)}$} in the reverse protocol, the entropy production is defined as  
\bb \label{eq:irr}
\Delta_\text{i}s[\gamma]=  \log(P_\text{d}[\gamma]/P_\text{r}[\gamma_\text{r}]). 
\ee
\noindent Averaged over all trajectories, Eq.\ref{eq:irr} verifies by construction the central Fluctuation Theorem $\left\langle e^{-\Delta_\text{i}s[\gamma]}\right\rangle_\gamma = \sum_{\gamma} P_\text{d}[\gamma] e^{-\Delta_\text{i}s[\gamma]} = 1$, and consequently the Second Law by convexity of the exponential $\left\langle \Delta_\text{i}s[\gamma]\right\rangle_\gamma \geq 0$. 
Such entropy production can always be split into two components involving a boundary term  $\Delta^\text{b}_\text{i}s[\gamma]$ and a conditional term $\Delta^\text{c}_\text{i}s[\gamma]$, with
\bb \label{eq:bound}
\Delta^\text{b}_\text{i}s[\gamma] = \log\left(\dfrac{p_\text{d}(\psi_0)}{p_\text{r}(\psi_N)}\right)
\ee 
\noindent and 
\bb \label{eq:cond}
\Delta^\text{c}_\text{i}s[\gamma] =\log\left(\dfrac{P_\text{d}[\gamma|\psi_0]}{P_\text{r}[\gamma_r|\psi_N]}\right).
\ee
\noindent We have introduced the conditional probability $P_\text{r}[\gamma_\mathrm{r}|\psi_N] = P_\text{r}[\gamma_\mathrm{r}]/p_\text{r}(\psi_N)$. 
As $|\langle \psi_{n+1} | {\cal G}(t_{n+1},t_{n})| \psi_n \rangle|^2 = |\langle \psi_n | {\cal G}^\text{r}(t_{n},t_{n+1})|\psi_{n+1} \rangle|^2$, the conditional term reduces to $0$. 
As stated above, $p_\text{d}(\psi_0)=1$ as the initial state is known, while $p_\mathrm{r}(\psi_N)= \pi_{{\cal K}_\gamma(t_N)}$. We get eventually $\Delta_\mathrm{i}s[\gamma] = -\log(\pi_{{\cal K}_\gamma(t_N)})$. The expression for the mean entropy production writes 
\bb \label{eq:VN}
\Delta_\mathrm{i} S = \langle \Delta_\text{i} s[\gamma] \rangle_\gamma = S_\text{VN}[\rho_\mathrm{s}(t_N)],
\ee

\noindent where $S_\text{VN}[\rho_\mathrm{s}(t_N)] = -\sum_{\cal K} \pi_{\cal K} \log(\pi_{\cal K} )$ is the Von Neumann entropy of the final mixed state \cite{Manzano-15}. By relating the change of the system's Von Neumann entropy to a well-defined, thermodynamic entropy production, Eq.\ref{eq:VN} allows quantifying the degree of irreversibility of a given measurement process. Contrary to the entropy produced during the relaxation in some heat bath, which can diverge as temperature approaches zero, here the measurement entropy is bounded by  $\Delta_\mathrm{i} S^\text{max} = -{\cal N}\ln {\cal N}$. From this study, it appears that quantum measurement is reversible, solely if the measurement process preserves the Von Neumann entropy of the system. If the outcomes are recorded, this means that measuring should not have any back-action on the system's pure state. If the outcomes are not recorded, measuring should not induce any decoherence in the system's state.   \\

\noindent \textbf{Thermodynamic quantities.} We now investigate the energetic implications of measurement induced irreversibility. To do so, we introduce the quantum analog to the system's internal energy, as the expectation value of the system's Hamiltonian if the system is in the pure quantum state $\ket{\psi_\gamma(t)}$
\bb \label{eq:U}
U_\gamma(t) = \bra{\psi_\gamma(t)} H_\mathrm{s}(t) \ket{\psi_\gamma(t)}.
\ee 
This quantity is generally understood as the average of some energy measurement performed on identical copies of the system. Energy fluctuations of quantum nature appear if the system is in a superposition of energy eigenstates. Consequently, the energy is said to be well defined, solely if the system's state is an energy eigenstate, to which the above definition is thus usually restricted \cite{Esposito-09,Campisi-RMP-11}. Here we extend this definition to any quantum state of the system's Hilbert space. It is still an operational quantity, as it can be fully reconstructed provided that the system's quantum trajectory and the applied Hamiltonian are known. In what follows, we use this quantity as our thermodynamic potential: As $U_\gamma(t)$ is homogeneous to an energy and characterizes the system alone, we shall simply call it internal energy. Let us underline that reconstructing quantities is often required in stochastic thermodynamics, e.g. heat and work exchanged during classical trajectories are inferred from the record of the system's evolution \cite{Berut-12}. Using Eq.\ref{eq:U}, the system's internal energy is defined for any quantum superposition of energy eigenstates, at any time of the quantum trajectory. In particular, it now takes a definite value after {\it and before} a quantum measurement, a mandatory condition to observe any energetic imprint related to the measurement process. 

``Heat'' and ``work'' are then defined by analogy with the classical situation: 
The work exchanged along the trajectory $W[\gamma]$ (resp. the heat $Q[\gamma]$) is identified with deterministic energy changes during the Hamiltonian evolution (resp. stochastic energy changes induced by measurement back-action):

\bb
W[\gamma] &=& \sum_{0\leq n\leq N-1} U_\gamma(t^{-}_{n+1}) -  U_\gamma(t^{+}_{n})\label{eq:W} \\
Q [\gamma] &=& \sum_{1\leq n\leq N} U_\gamma(t^{+}_{n}) -  U_\gamma(t^{-}_{n})\label{eq:Q}.
\ee

\noindent $t_n^+$ (resp. $t_n^-$) stands for the time immediately after (resp. before) {the time $t_n$ of the measurement}. Work quantifies the energy exchanged with the driving source and vanishes if the Hamiltonian is time-independent. The elementary work performed during time $dt$ corresponds to a system's Hamiltonian variation $dH_\mathrm{s}(t)$ and writes:
\bb
\delta W_\gamma(t) = \bra{\psi_\gamma(t)} dH_\mathrm{s}(t) \ket{\psi_\gamma(t)}.\label{dWI}
\ee
On the other hand, ``heat'' solely appears if the measurement process induces some back-action on the system's state, such that $\ket{\Psi_\gamma(t_n^-)} = {\cal G}(t_n, t_{n-1})\ket{\psi_n}$ differs from $\ket{\Psi_\gamma(t_n^+)} = \ket{m_{{\cal K}_\gamma(t_n)}}$. Note that this quantity has no classical equivalent: It is solely due to genuinely quantum energy fluctuations, which can take place at zero temperature. Therefore we shall simply call it ``quantum heat'', denoted in the following $Q_\text{q}[\gamma]$. 
This quantity already appears in \cite{Manzano-15}, its physical origin being qualified as ``obscure".  Actually, each situation can give rise to a detailed energy balance accounting for the quantum heat contribution (see the Supplemental for an example of such analysis). But more simply, the quantum heat appears as a natural byproduct of the standard quantum formalism, as soon as Eq.\ref{eq:U} is used. In this sense, it can be seen as a straightforward thermodynamic consequence of the measurement postulate, by which a quantum system coupled to a measuring device is actually an open system. In what follows, we shall therefore not search for further microscopic justifications of the quantum heat, but rather treat it as an energetic imprint of measurement induced back-action, and a key concept to reveal genuine quantum features. \\

\begin{figure}[t]
\begin{center}
\includegraphics[scale=0.3]{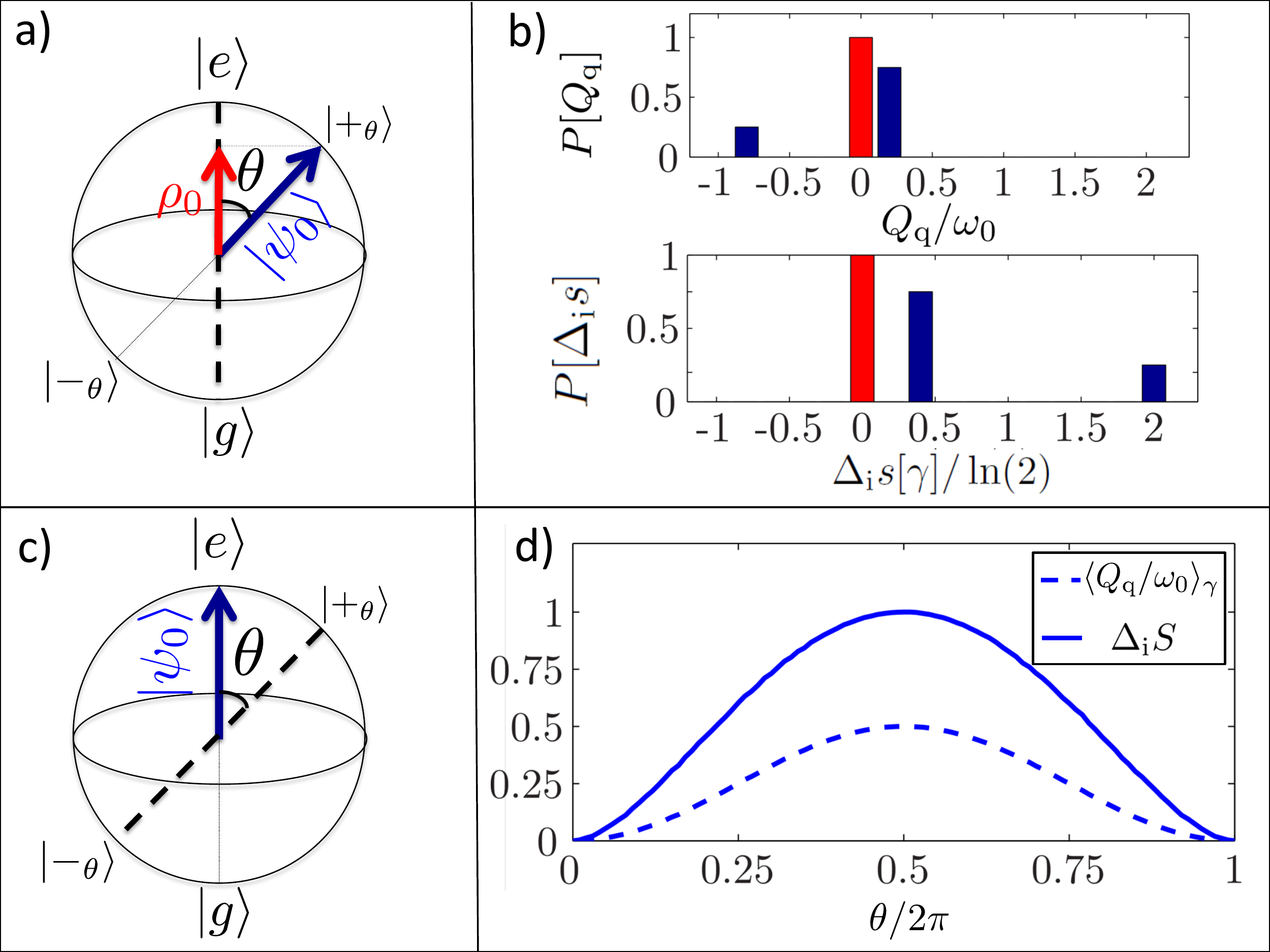}
\end{center}
\caption{Distributions of quantum heat $P[Q_\mathrm{q}]$ for simple protocols defined by the preparation and measurement of a Qubit in arbitrary bases $\{ \ket{+_\theta} ; \ket{-_\theta} \}$, with $\ket{+_\theta} = \cos(\theta/2)\ket{e} + \sin(\theta/2)\ket{g}$ and $ \ket{-_\theta} = -\sin(\theta/2)\ket{e} + \cos(\theta/2)\ket{g}$. $\ket{e}$ and $\ket{g}$ are the excited and ground states of the Qubit of respective internal energies $U_\mathrm{e} = \hbar \omega_0/2$, and $U_\mathrm{g} = -\hbar \omega_0/2$. a), b): Energetic and entropic signatures of quantum measurement. a): The Qubit is either prepared in the pure state $\ket{\psi_0} = \ket{+_\theta}$ of internal energy $U_{+_\theta} = \frac{\hbar \omega_0}{2} \cos^2(\theta)$ (dark blue arrow) or picked from a mixture $\rho_0$ of the states $\ket{e}$ and $\ket{g}$ with respective probabilities $\cos^2(\theta/2)$ and $\sin^2(\theta/2)$ (red arrow). The Qubit is then measured in the $\{\ket{e}; \ket{g}\}$ basis.  b): Normalized  distribution of entropy production $P[\Delta_\text{i} s]$ and of quantum heat $P[Q_\mathrm{q}]$ for $\theta = \pi/3$. $P[Q_\mathrm{q}]$ has non zero components if the initial state is $\ket{\psi_0}$, corresponding to $U_\mathrm{e}-U_{+_\theta}=\sin^2(\theta/2)$ (if the Qubit is measured in $\ket{e}$), and to $U_\mathrm{g}-U_{+_\theta}=-\cos^2(\theta/2)$ (if the Qubit is measured in $\ket{g}$). In the same case, entropy production distribution also features two peaks. $P[Q_\text{q}]$ and $P[\Delta_\text{i} s]$ are delta-peaked at  zero (red bar) if the initial state is randomly picked from $\rho_0$. In both cases, the average quantum heat $\langle Q_\mathrm{q} \rangle_\gamma$ is zero. c): The Qubit is prepared in the pure state $\ket{\psi_0} = \ket{e}$ (dark blue arrow) and measured in the $\{\ket{+_\theta},\ket{-_\theta}\}$ basis. d) Average quantum heat $\langle Q_\mathrm{q}\rangle_\gamma$ (solid blue) and average entropy production  $\langle \Delta_\text{i} s\rangle_\gamma$ as a function of $\theta$.  The mean quantum heat is null, solely if the measurement and Hamiltonian eigenbases commute ($\theta = 0$)}\label{f:fig2}
\end{figure}

\label{p:Heat-q}
\noindent \textbf{Properties of quantum heat.} 
For the sake of clarity we consider the case where the driving source is switched off, such that the system's Hamiltonian  $H_\text{s}$ is constant and no work is exchanged. A two-points trajectory $\gamma$ is defined by preparing the system in the arbitrary state $\ket{\psi_0}$ of internal energy $U_0$ and measuring it in the basis $\{ \ket{m_{\cal K}} \}$. During the measurement, the system eventually jumps to the state $\ket{m_{{\cal K}_\gamma}}$ of energy $U_{{\cal K}_\gamma}$: Provided that the eigenstates $\{ \ket{m_{\cal K}} \}$ have different internal energies, a jump can give rise to different quantum heat contributions $Q_\mathrm{q}[\gamma] = U_{{\cal K}_\gamma}-U_0$. Repeating this experiment many times allows reconstructing a normalized distribution of quantum heat $P[Q_\mathrm{q}] = \sum_\gamma P_\mathrm{d}[\gamma] \delta^{\cal D}(Q_\mathrm{q}-Q_\mathrm{q}[\gamma])$ where $\delta^{\cal D}$ stands for the Dirac distribution. The example of a Qubit is pictured in Fig.\ref{f:fig2}. The distributions of quantum heat $P[Q_\text{q}]$ and entropy production $P[\Delta_\text{i}s] = \sum_\gamma P_\text{d}[\gamma]\delta^{\cal D}(\Delta_\text{i} s - \Delta_\text{i} s[\gamma])$ have non zero components, solely if $\ket{\psi_0}$ has coherences in the basis $\{ \ket{m_{\cal K}} \}$, such that quantum measurement has some finite back-action on the system's state. On the contrary, both distributions are delta-peaked at zero if $\ket{\psi_0}$ is randomly picked from a statistical ensemble of eigenstates $\{ \ket{m_{\cal K}} \}$. Non zero components in the distributions of quantum heat and entropy production are therefore contemporary phenomena, that can be seen as the energetic and entropic signatures of measurement induced irreversibility. Formal relations between the two distributions can be drawn, e.g. in the case of the quantum Jarzynski's equality (See Supplemental).

If the measured observable ${\cal M}$ commutes with the system's Hamiltonian, we obviously have $\langle U_{{\cal K}_\gamma} \rangle_\gamma = U_0$, such as the average quantum heat vanishes: $\langle Q_\mathrm{q}[\gamma] \rangle_{\gamma} = \sum_\gamma P_\mathrm{d}[\gamma] Q_\mathrm{q}[\gamma]=0$. On the contrary, if ${\cal M}$ and $H$ do not commute, performing a measurement changes the system's average internal energy (Fig.\ref{f:fig2}d). Remarkably here, measuring can provide energy to the system, strengthening the analogy between a heat bath and a measuring apparatus. Such mechanism can be further exploited, e.g. to develop genuinely quantum engines, driven by quantum measurement \cite{zeno}.\\

\label{p:OpenSyst}
\noindent \textbf{Generalized measurements and quantum open systems.} The framework presented above can be extended to generalized measurements, such as weak or destructive measurements. Here the number of measurement outcomes is not bounded by the dimension of the system's Hilbert space, in particular their set can be continuous. Recording the outcome $m_{\cal K}$ results in applying the so-called Kraus operator $M_{\cal K}$ on the system's quantum state $\ket{\psi}$ \cite{Raimond-06}. This event occurs with a probability $p_{\cal K} = \bra{\psi} M_{\cal K}^\dagger M_{\cal K} \ket{\psi}$, and the set of Kraus operators $\{ M_{\cal K} \}$ verifies {$\sum_{\cal K} M_{\cal K}^\dagger M_{\cal K}=1$}.  If the outcomes are not read, the system's state becomes mixed, and its evolution under the measurement process is described by the completely positive trace-preserving map $\rho_\text{s} \rightarrow \sum_{\cal K} M_{\cal K} \rho_\text{s} M^\dagger_{\cal K}$. 

Generalized measurements provide fruitful insights into the physics of quantum open systems. Such systems are at the core of most physical situations currently studied in quantum thermodynamics, e.g. the coupling of a quantum system to a thermal bath \cite{Esposito-09,Campisi-RMP-11}, or the continuous measurement of a Qubit \cite{Alonso-16}, on which we shall focus from now on. The most general equation describing the average evolution of such open system is the Lindblad master equation

\bb \label{Meq}
\dot{\rho_\mathrm{s}}(t) = -i[H_\mathrm{s}(t),\rho_\mathrm{s}]+ {\cal L}[\rho_\mathrm{s}].
\ee

\noindent We have introduced the Lindbladian superoperator ${\cal L}[\rho]=\sum_{k=1}^{{\cal N}_\text{s}^2-1}  \Gamma_k(L_k \rho_\mathrm{s} L^\dagger_k - \frac{1}{2} \{ \rho_\text{s}, L^\dagger_k L_k \})$ expressed in term of at most ${\cal N}_\text{s}^2-1$ Lindblad operators $\{L_k\}$ acting on the system's Hilbert space and rates $\Gamma_k$.  Because of the driving source, the Lindbladian may also depend on time \cite{Gasparinetti-13,Elouard-15,Elouard-16}, but we shall not systematically write this time-dependence in the following. It is always possible to rewrite the evolution generated by Eq.\ref{Meq} as some generalized measurement performed on the quantum system at each time step $dt$, that involves a set of Kraus operators $\{ M_{\cal K} \}$:
\bb \label{Kraus}
\rho_\mathrm{s}(t + dt) = \sum_{{\cal K}} M_{\cal K} \rho_\mathrm{s}(t) M^\dagger_{\cal K} 
\ee
\noindent This process can be interpreted as the result of some unrecorded, continuous measurement performed on a Markovian reservoir weakly coupled to the system, with stochastic outcomes $m_{\cal K}$  (Fig.\ref{fig1}c). The choice of a given detection scheme unambiguously fixes the set of Kraus operators and the so-called unraveling of the master equation. 
For a Quantum Jump (QJ) unraveling, the set of measurement outcomes is discrete and bounded to ${\cal N}^2_\text{s}$. In this case the set of Kraus operators consists of at most ${\cal N}^2_\text{s}-1$ ``jump'' operators, corresponding to a macroscopic back-action on the system's state, i.e., a quantum jump. Each jump is described by one of the Lindblad operator $L_k$ and verifies:
\bb
M_k = \sqrt{\Gamma_k dt}L_k
\ee
\noindent The set of Kraus operators also contains a ``no-jump" operator ensuring trace conservation $M_0(t) = \idop-idt H_\text{eff}(t)$, where $H_\text{eff}(t) = H_\text{s}(t) - (dt/2)\sum_{k=1}^{{\cal N}_\text{s}^2-1}\Gamma_k L_k^\dagger L_k$ is the effective (non-hermitian) system's Hamiltonian. 

Reciprocally, the readout of some continuous observable of the reservoir corresponds to the quantum state diffusion (QSD) unraveling. In this case, the set of Kraus operators is also continuous: The record of some outcome between ${\cal K}$ and ${\cal K}+d{\cal K}$ gives thus rise to some infinitesimal back-action $M_{\cal K}$ on the system's state: \cite{Brun-00,Breuer-02} 
\bb
M_{{\cal K}} (t) &=& \bigg(\idop-idtH_\text{eff}(t)
 + \sum_{k=1}^{{\cal N}_\text{s}^2-1}\sqrt{\Gamma_k} dw^{\cal{K}}_k(t)L_k\bigg)\nonumber\\ 
&&\times\prod_k \sqrt{p(dw^{\cal K}_k(t))}.
\ee 
\noindent In this situation all the jumps described by $L_k$ take place with a weight determined by the so-called Wiener increment $dw^{\cal K}_k(t)$, whose distribution over all possible trajectories verifies
\bb
\left\langle dw^{\cal K}_k(t) \right\rangle_\gamma &=& 0\label{dw_mean}\\
\left\langle {dw^{\cal K}_k(t)}^* dw^{\cal K}_l(t')\right\rangle_\gamma &=& dt \delta^{\cal D}(t-t')\delta_{kl}.\label{dw_var}
\ee

\noindent Whatever the unraveling, knowing the initial state $\ket{\psi_0}$ of the system and the complete measurement record $m_{{\cal K}_\gamma(t)}$ allows the full reconstruction of the system's quantum trajectory $\ket{\psi_\gamma(t)}$: The state $\ket{\psi_\gamma(t+dt)}$ is obtained by applying the operator $M_{{\cal K}_\gamma(t)}$ to the system's state, and then renormalizing. The evolution of $\ket{\psi_\gamma(t)}$ can be formulated in terms of a stochastic Schr\"odinger equation (see Methods). Again the density matrix solution of Eq.\ref{Meq} is recovered by averaging over the trajectories: $\rho_\mathrm{s}(t) = \langle \ket{\psi_\gamma(t)} \bra{\psi_\gamma(t)}\rangle_\gamma$.
Such interpretation is the historical way quantum trajectories were introduced \cite{Dalibard-92}. Remarkably in this picture, the reservoir is seen as part of some monitoring channel extracting information on the system, and the stochasticity in the system's evolution primarily comes from quantum measurement, just like in the ideal situation presented above. Owing to impressive progresses in detection efficiencies, experimental reconstruction of quantum trajectories is nowadays state of the art, as demonstrated by series of pioneering results obtained with trapped ions \cite{Barreiro-11}, and later on in Cavity Quantum ElectroDynamics \cite{Gleyzes-07} and circuit QED \cite{Riste-13,Murch-13,Weber-14,Roch-14}.\\

\noindent \textbf{Thermodynamic quantities for quantum open systems.}
The system's internal energy $U_\gamma(t)$ and work increment  $\delta W_\gamma(t)$ during time $dt$ are still defined by Eq.\ref{eq:U} and \ref{dWI} respectively, while the heat increment is defined as $\delta Q_\gamma(t) = dU_\gamma(t) - \delta W_\gamma(t)$. Both the work and the heat increment can be expressed as the expectation value of some work/heat operator, taken in the system's quantum state $\ket{\psi_\gamma(t)} =  \sum_{i=1}^{{\cal N}_\text{s}} \psi_i(t) \ket{i}$ (See Methods). The choice of a specific basis ${\cal B} = \{ \ket{i} \}$ to write $\ket{\psi_\gamma(t)}$ allows splitting the heat and work increments into a classical and a quantum contribution, the later vanishing if the coherences $\psi_i^* \psi_j$ of the system's state are zero.

While the work component still quantifies the energy exchanged with the driving source, the heat contribution now involves two mechanisms by which the system's energy can change: The deterministic non-Hermitian evolution and the stochastic quantum jumps. These two mechanisms catch an essential difference with respect to classical thermodynamics, namely, that the reservoir plays a double role: It does not only exchange energy with the system like a regular bath, but it also extracts information on the system's state, erasing its coherences like a measuring apparatus. In this spirit, ${\cal B}$ can always be chosen, such that the classical and quantum heat respectively reflect the classical and quantum energy fluctuations induced by the reservoir.

To fix the ideas, let us consider a QJ unraveling, where each jump operator $M_{{\cal K}}$ (${\cal K}\geq 1$) is defined as
\bb \label{eq:QJ}
M_{\cal K} = \sqrt{\Gamma_{\cal K}dt} \ket{j({\cal K})}\bra{i({\cal K})}.
\ee

\noindent The set of orthogonal states $\{ \ket{i({\cal K})} \}$ of respective internal energies $\epsilon_{i({\cal K})}$ defines a natural basis ${\cal B}$. Each jump consists in a projection on the state $\ket{i({\cal K})}$, followed by the transition $\ket{i({\cal K})} \rightarrow \ket{j({\cal K})}$, such that the whole process can be seen as a destructive measurement of the state $\ket{i({\cal K})}$. In this spirit, let us consider the pure initial state $\ket{\psi_0}$ of internal energy $U_0$. The jump ${\cal K}$ occurs with a probability {$\Gamma_{\cal K}dt|\langle i({\cal K}) \ket{\psi_0}|^2$} and gives rise to a total heat exchange $Q = \epsilon_{j({\cal K})} -U_0$. This heat exchange can always be rewritten as $Q_\text{cl}({\cal K}) +  Q_\text{q}$, where $Q_\text{q} = \epsilon_{i({\cal K})} - U_0$ is the quantum energy fluctuation induced by the projection on the state $\ket{i({\cal K})}$, and $Q_\text{cl}({\cal K}) = \epsilon_{j({\cal K})}  - \epsilon_{i({\cal K})} $ is usually interpreted as the classical energy exchange with the reservoir. The value of $Q_\text{cl}({\cal K})$ is fixed by the jump, while $Q_\text{q}$ can take any value. Just like in the ideal situation, the quantum heat solely shows up if the initial state $\ket{\psi_0}$ carries coherences in the basis ${\cal B}$.\\

\noindent {\textbf{Irreversibility of quantum trajectories.} The entropy produced by the trajectory $\gamma$ is still defined by Eq.\ref{eq:irr}, where the reverse protocol consists in applying the time-reversed Hamiltonian $H_\text{s}(t_N-t)$ while the reservoir is continuously monitored. The conditional terms $P_\mathrm{d}[\gamma|\psi_0]$ and $P_\mathrm{r}[\gamma_\mathrm{r}|\psi_N]$ now equal the probability of the sequence of jumps corresponding to the direct/reverse protocol respectively \cite{Brun-00}:  
 
\bb \label{eq:ptraj}
P_\mathrm{d}[\gamma|\psi_0] = | \bra{\psi_N} \left(\prod_{n=0}^N M_{{\cal K}_\gamma(t_n)}\right)\ket{\psi_0} |^2.
\ee

\bb \label{eq:pr}
P_\mathrm{r}[\gamma_\mathrm{r}|\psi_N] = |\bra{\psi_0} \left(\prod_{n=0}^N M^\text{r}_{{\cal K}_\gamma(t_{N-n})}\right)\ket{\psi_N} |^2.
\ee
We have discretized the time interval and defined the times $t_n = t_0 + ndt$ at which the measurements are performed on the reservoir. $\ket{\psi_N}$ stands for the final state of the direct trajectory. So far quantum fluctuation theorems have systematically involved protocols including some final projective measurement \cite{Manzano-15,Esposito-09}, where $\ket{\psi_N}$ is the corresponding eigenstate. In the present framework, the considered trajectories do not necessarily end up with such a projective measurement: Then $\ket{\psi_N}$ is the system's state after the final quantum jump induced by the reservoir.

We have introduced the time-reversed Kraus operators $M^\text{r}_{\cal K}$. Their definition is well established in the case of thermal fluctuations induced by some heat bath of inverse temperature $\beta$. This situation is usually described by a QJ unraveling.
Introducing the thermal equilibrium state $\pi_\text{s} = Z^{-1}\exp(-\beta H_\text{s})$ where $Z$ is the canonical partition function, the time-reversed Kraus operators giving rise to a jump (${\cal K}\geq 1$) write \cite{Crooks-08}
\bb
M^\text{r}_{\cal K}=\sqrt{\pi_\mathrm{s}} M^\dagger_{\cal K} \sqrt{\pi_\mathrm{s}^{-1}},\quad {\cal K}\geq 1,\label{MrQJ}
\ee
which simplifies into
\bb
M^\text{r}_{\cal K}=e^{\frac{\beta Q_\text{cl}({\cal K})}{2}} M^\dagger_{\cal K}.\label{Mrbath}
\ee
\noindent The time-reversed Kraus ``no-jump'' operator verifies 
\bb
M_0^\text{r} (t_n) = 1+idt H_\text{eff}^\dagger(t_n).
\ee 
\noindent These expressions guarantee that if the system is at equilibrium, a sequence of two consecutive jumps in the direct protocol occurs with the same probability as the reverse sequence of time-reversed jumps in the reverse protocol. If the temperature is finite, this set of time-reversed operators allows deriving the quantum Jarzynski's equality for quantum open systems (see Supplemental). In what follows we rather focus on the less investigated zero temperature case: In this case only the jumps giving rise to a negative classical heat contribution can occur, and lead to a divergence of entropy production. This characterizes for instance the spontaneous emission of a Qubit (See below). 

Eq.\ref{MrQJ} can also be used to derive the time-reversed operators corresponding to the continuous monitoring of some system's observable $X$, in the absence of any thermal bath. Here $\pi_\text{s}$ is a state invariant under the measurement process, verifying $[\pi_\text{s},X]=0$. Such continuous measurement is described by a QSD unraveling involving a single Lindblad operator $X = L_1$ characterized by the Wiener increment $dw(t) = dw_1(t)$ \cite{Jacobs-06}. The direct and reverse Kraus operators corresponding to the outcome ${\cal K}$ write

\begin{equation}
M_{\cal K} = [1+idtH_\text{eff}(t) + \sqrt{\Gamma^*}dw^{\cal K}(t) X]\sqrt{p(dw^{\cal K}(t))},
\end{equation}
\noindent and 
\begin{equation}
M_{\cal K}^\text{r} = [1+idtH_\text{eff}^\dagger (t)+ \sqrt{\Gamma^*}\left({dw^{\cal K}(t)}\right)^* X] \sqrt{p(dw^{\cal K}(t))}.
\end{equation}

\noindent We have introduced the decoherence rate $\Gamma^*$, which quantifies the strength of the measurement per unit of time. These expressions allow in particular quantifying the irreversible character of a continuous measurement. Just like in the case of projective measurements, the conditional term $\Delta_\text{i}^\text{c}s[\gamma]$ appearing in Eq.\ref{eq:irr} vanishes. Eventually the mean entropy production writes 
\bb
\langle \Delta_\text{i}s[\gamma]\rangle_\gamma = S_\text{VN}[\rho_\text{s}[t_N]]- S_\text{VN}[\rho_\text{s}[t_0]].
\ee 

\noindent We now exploit our formalism to analyze two typical situations of quantum optics: The spontaneous emission of a Qubit prepared in a coherent superposition of energy eigenstates, and the stabilization of a quantum state by feedback protocol.\\

\noindent \textbf{Thermodynamics of spontaneous emission.} At the initial time $t_0$, a Qubit of respective ground and excited states $\ket{g}$ and $\ket{e}$ with transition frequency $\omega_0$ is prepared in the quantum superposition $\ket{+_x} = (\ket{e} + \ket{g})/\sqrt{2}$. The Qubit is coupled to a zero temperature reservoir monitored with a photo-counter. This corresponds to a QJ unraveling involving a single jump operator $M_{1} = \sqrt{\Gamma dt} \sigma_-$, where $\sigma_-=\ket{g}\bra{e}$ is the lowering operator and $\Gamma$ the spontaneous emission rate \cite{Raimond-06}. The effective Hamiltonian writes $H_\mathrm{eff} = \hbar (\omega_0  - i \Gamma/2) \sigma_+ \sigma_-$, where $\sigma_+ = \sigma_-^\dagger$. The Qubit's trajectories are computed between $t_0=0$ and $t_N = t$, the typical monitoring time $dt$ verifying $dt\ll \Gamma^{-1}$.

\begin{figure}[t]
\begin{center}
\includegraphics[scale=0.44]{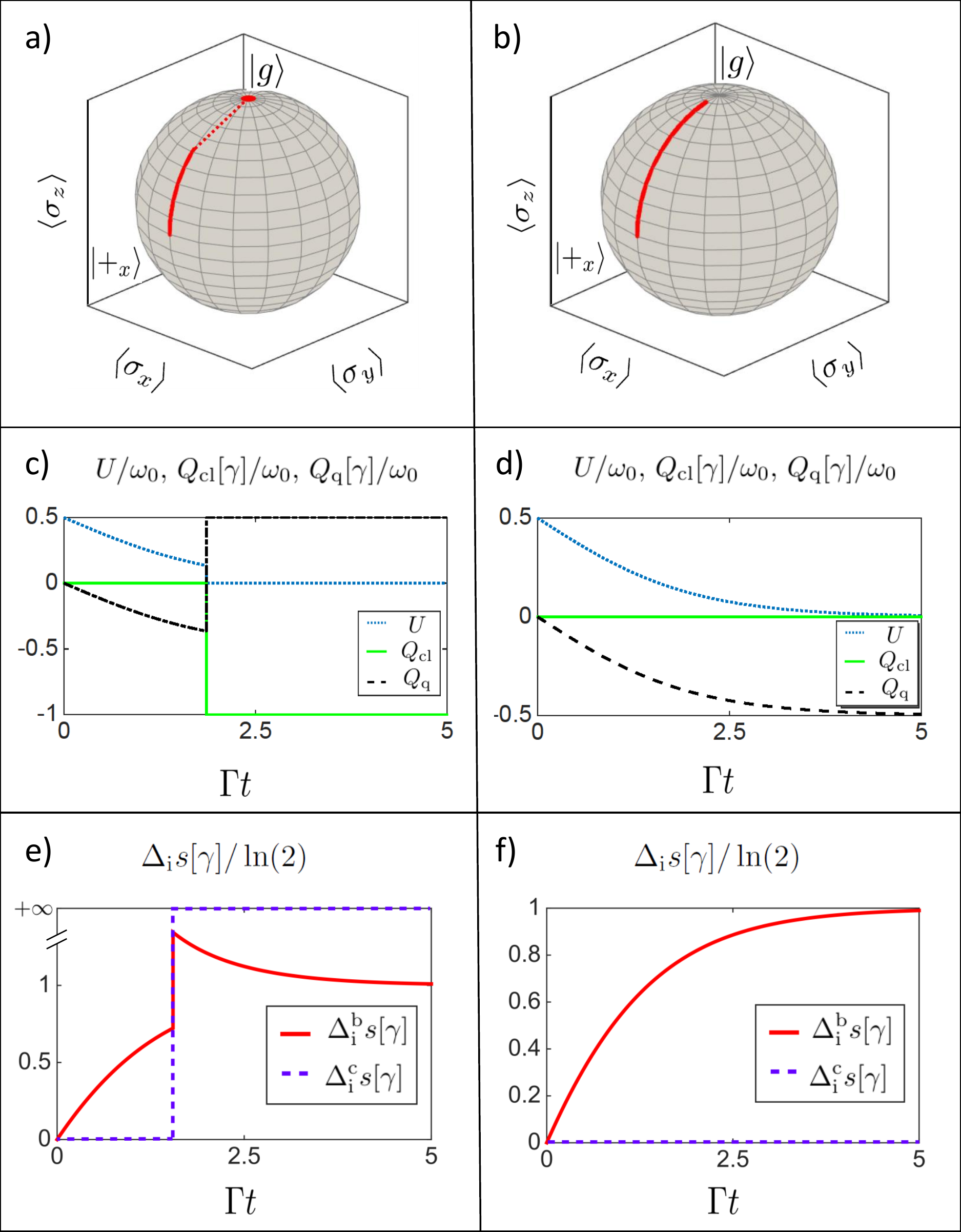}
\end{center}
\caption{Thermodynamics of spontaneous emission. A Qubit is prepared in the $\ket{+_x}$ state and coupled to a zero-temperature reservoir monitored with a photo-counter. a), b): Quantum trajectories of the Qubit's state in the Bloch sphere, in the case of a jump trajectory (a) and no-jump trajectory (b). c), d): Internal energy, classical and quantum heat contributions in units of $\hbar \omega_0$ as a function of time. e),f): two contributions to entropy production: $\Delta_\text{i}^\text{c}s[\gamma]$ and $\Delta_\text{i}^\text{b}s[\gamma]$, in unit of bits.}
\label{fig3} \end{figure}

There are two classes of trajectories giving rise to two possible final states $\ket{\psi_N}$. In the ``jump" trajectories, the Qubit relaxes to the ground state by emitting a photon, such that $\ket{\psi_N} = \ket{g}$, which happens with a probability $p_\text{j}(t) = (1-e^{-\Gamma t})/2$. Reciprocally the ``no-jump" trajectory occurs with the probability $p_\text{nj}(t) = (1+e^{-\Gamma t})/2$: The Qubit deterministically evolves under $H_\text{eff}$ until $t=t_N$, such that the final state writes $\ket{\psi_N} = \ket{\psi_\text{nj}(t)}=\left( e^{-(\Gamma+i\omega_0) t/2} \ket{e}+ e^{i\omega_0 t/2}\ket{g}\right) /{\sqrt{1+e^{-\Gamma t}}}$. Remarkably for $t\gg \Gamma^{-1}$, the Qubit also ends up in the ground state, while no photon has been emitted: Just like recording a click, not detecting a photon increases the knowledge on the system's state. Eventually, the whole process of spontaneous emission at large times $t\gg \Gamma^{-1}$ can be seen as a measurement of the Qubit in its energy basis, recording  (resp. not recording) a click boiling down to measuring the Qubit in $\ket{e}$ (resp. $\ket{g}$). Such measurement process is destructive as the Qubit always ends up in $\ket{g}$. Remarkably at finite times the measurement is performed between two non-orthogonal states, i.e. $\ket{e}$ (if a click is recorded) and $\ket{\psi_\text{nj}(t)}$ (if no click is recorded).

We have analyzed the energetic and entropic signatures of this non-ideal measurement process (See Methods and Fig.\ref{fig3}). The jump induces an exchange of classical heat $Q_\text{cl} = -\hbar \omega_0$, which remains null for the no-jump trajectory. After the jump, the quantum heat equals $Q_\text{q} = \hbar \omega_0/2$, which is consistent with the Qubit being measured in $\ket{e}$. Reciprocally along the no-jump trajectory, $Q_\text{q}$ converges towards $-\hbar \omega_0/2$ for $t\gg \Gamma^{-1}$, which corresponds to a measurement in $\ket{g}$. Entropy production is computed using Eqs.\ref{eq:irr}, \ref{eq:bound} and \ref{eq:cond}. As long as no jump takes place, the conditional term $\Delta_\text{i}^\text{c}s[\gamma](t)$ defined by Eq.\ref{eq:cond} equals $0$ as the evolution is deterministic. It diverges as soon as a click is recorded, which is typical of spontaneous emission as argued above. 

Interestingly, the boundary term remains finite in any case, verifying $\Delta_\text{i}^\text{b}s[\gamma](t) = -\log[p_\text{nj}(t)]$ (resp $-\log[p_\text{j}(t)]$) for the no-jump trajectory (resp. for any jump trajectory). The average boundary term for entropy creation therefore writes $\Delta_\text{i}^\text{b}S(t) = \langle \Delta_\text{i}^\text{b}s[\gamma] \rangle_\gamma (t) = {\cal H}[p_\text{nj}(t)]$, where $ {\cal H}[p] = -p\log(p ) - (1-p)\log(1-p)$ stands for the Shannon's entropy. This entropy quantifies the information acquired at time $t$, i.e. the measurement into one of the two non-orthogonal states $\ket{e}$ or $\ket{\psi_\text{nj}(t)}$. For $t\gg \Gamma^{-1}$, $\ket{\psi_\text{nj}(t)}$ converges towards $\ket{g}$ and the two states become orthogonal, such that $\Delta_\text{i}^\text{b}S$ converges towards $S_\text{VN}[\rho_\text{s}(t)] = \log(2)$.\\

\label{p:ExFeedback}
\noindent \textbf{Work cost of a feedback protocol.} We now investigate the energetic requirements to perform a state-stabilization protocol. A Qubit prepared in $\ket{+_x}$ is weakly monitored in the $\{\ket{e},\ket{g} \}$ basis, giving rise to a continuous, stochastic measurement record $y_\gamma(t)$. Under the monitoring, the system's state evolves by infinitesimal quantum jumps towards $\ket{e}$ or $\ket{g}$ within a typical decoherence time ${\Gamma^{*}}^{-1}$. Here the coupling to a thermal bath is neglected and we solely consider the stochastic perturbation induced by the decoherence process. This situation is described by a QSD unraveling involving a single Lindblad operator $L_z = \sqrt{\Gamma^*}\sigma_z$ \cite{Wiseman-09}. Introducing the Wiener increment $dw_\gamma(t)=y_\gamma(t)\sqrt{4\Gamma^*}dt$ \cite{Jacobs-06}, the system's evolution is ruled by the stochastic Schr\"odinger equation : 

\begin{align}
d\ket{\psi_\gamma} =\bigg[ -i\dfrac{\omega_0 dt}{2}\sigma_z - \dfrac{1}{2}\Gamma^* dt\left(\sigma_z-\left\langle\sigma_z\right\rangle\right)^2\nonumber\\
 +\sqrt{\Gamma^*}dw_\gamma(t)(\sigma_z-\left\langle\sigma_z\right\rangle)\bigg]\ket{\psi_\gamma}.\label{M:SSEdecoh}
\end{align}

\begin{figure}[t]
\begin{center}
\includegraphics[scale=0.33]{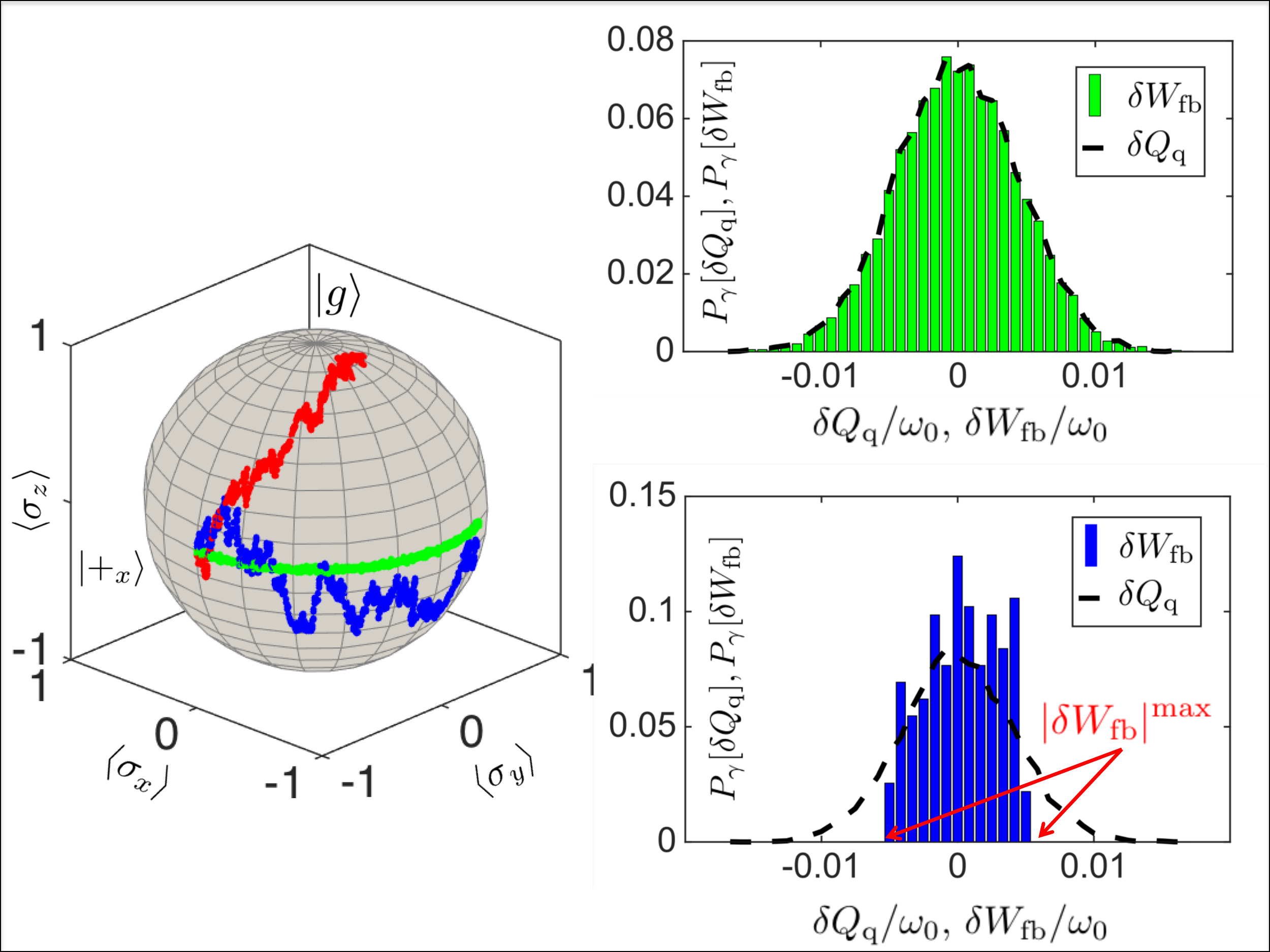}
\end{center}
\caption{Analysis of a feedback protocol stabilizing the state {$\ket{\psi_\text{target}(t)} = \exp(-i\omega_0 \sigma_z t/2)\ket{+_x}$}. Left: Trajectory in the Bloch sphere in the case of perfect feedback (green), imperfect feedback (blue) and without feedback (red). Right: Normalized distributions of quantum heat increments $P_\gamma[\delta Q_\text{q}]$ (dashed black) and feedback work $P_\gamma[\delta W_\text{fb}]$ {(bars)} performed by the feedback source ${\cal F}$. The work distribution $P_\gamma[\delta W_\text{fb}]$ is defined like in Eq.\ref{eq:dist}. Top right: The two distributions match, such that the state is perfectly stabilized. Bottom right: The distribution of $P_\gamma[\delta W_\text{fb}]$ is bounded and the feedback is not perfect.  {\it Parameters:} Evolution time $T= 1.5 / \omega_0$, pure dephasing rate $\Gamma^* = 0.1 \omega_0${, feedback work cutoff: $\vert\delta W_\text{fb}\vert^\text{max} = 0.05 \hbar \omega_0$}.}
\label{fig4} \end{figure}

\begin{figure}[t]
\begin{center}
\includegraphics[scale=0.4]{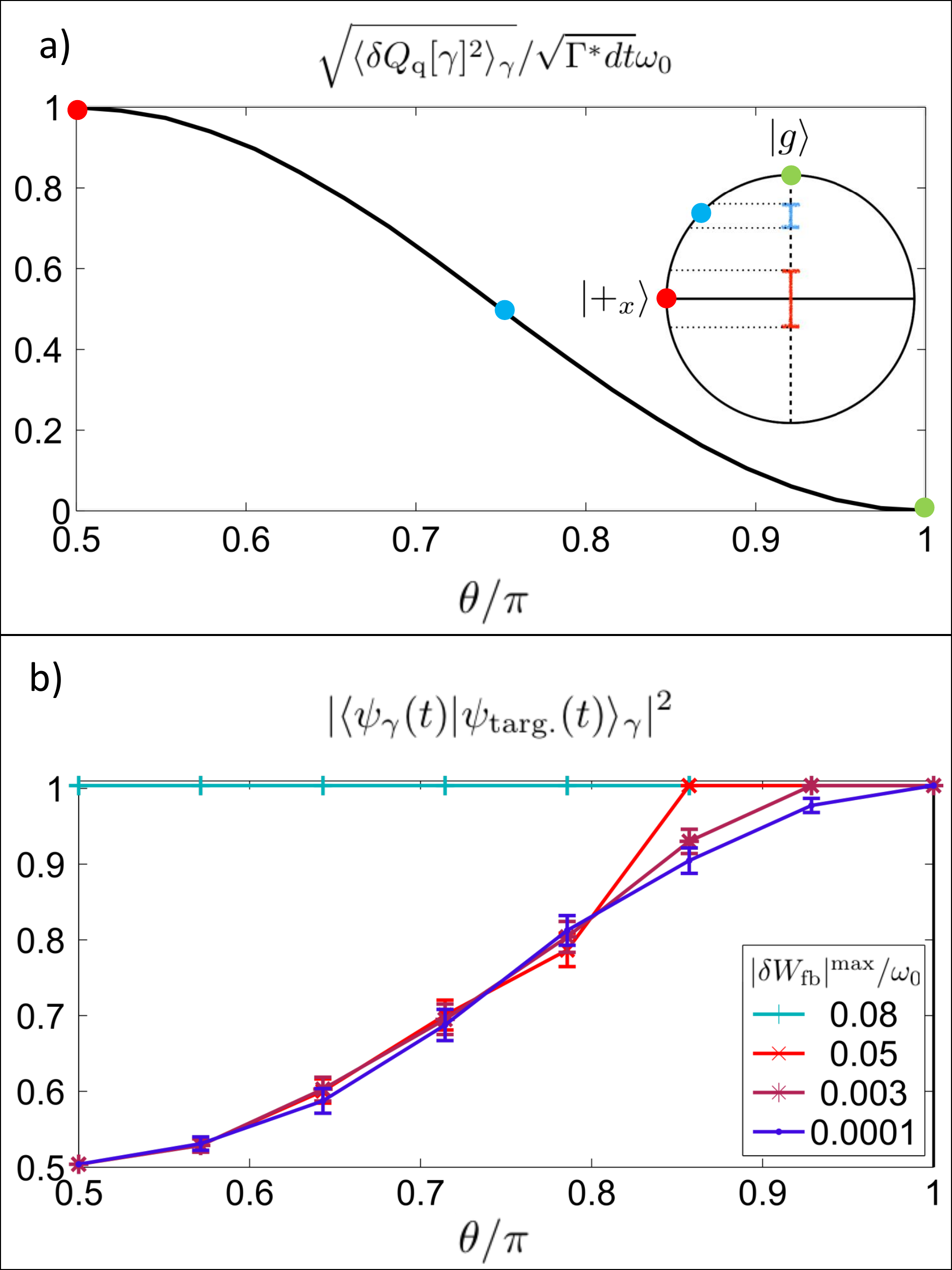}
\end{center}
\caption{
{Efficiency of the feedback protocol depending on the state to stabilize $\ket{\psi_\text{targ.}}$. a): Standard deviation of the quantum heat increment $\delta Q_\text{q}[\gamma](t)$ depending on the latitude $\theta$ on the Bloch sphere of the state $\ket{\psi_\text{targ.}(t)}$. b) : Average fidelity of the final state $\ket{\psi_\gamma(t)}$ to the target state $\ket{\psi_\text{targ.}(t)}$ as a function of the latitude $\theta$, for different values of the feedback work cutoff $|\delta W_\text{fb}|^\text{max}$. The error bars stand for the $99\%$ confidence interval. The trajectories last for $t= 20/\Gamma^*$.}
}
\label{fig5} \end{figure}

\noindent During a time step, the increment of internal energy writes (See Methods)
\bb \label{eq:fluc}
dU_\gamma(t) = \delta Q_\mathrm{q}[\gamma](t) = 4\sqrt{\Gamma^*}\hbar \omega_0 dw_\gamma(t)\vert\left\langle\sigma_- \right\rangle\vert^2.
\ee
\noindent Such energy fluctuations are all the larger as the measurement strength $\Gamma^*$ and the Qubit's coherences $\vert\left\langle\sigma_- \right\rangle\vert$ are large. They vanish when the Qubit's state has converged into one of the stable points $\ket{e}$ or $\ket{g}$ where measurement is completed. Once integrated between $0$ and $t \gg {\Gamma^{*}}^{-1}$, the quantum heat and internal energy change converge towards $\Delta U[\gamma] = Q_\text{q}[\gamma] = \pm \hbar \omega_0/2$ as the Qubit's state is approaching $\ket{e}$ or $\ket{g}$.

In order to counteract decoherence, the measurement record is continuously sent to a feedback source ${\cal F}$, which subsequently stabilizes the state $\ket{\psi_+(t)} = e^{-i\omega_0 t \sigma_z/2} \ket{+_x}$ by driving the Qubit with the Hamiltonian $H_\text{fb}=g(t) \sigma_y$. We assume the response time $\tau_\mathrm{fb}$ of the feedback loop is much shorter than the typical monitoring time, $\tau_\mathrm{fb} \ll dt$. We have studied the trajectories followed by the Qubit under monitoring and feedback, between the initial time $t_0 = 0$ and $t_N = t$. Each jump at time $t_k$ gives rise to a quantum heat increment $\delta Q_\mathrm{q}[\gamma](t_k)$, which must be compensated by some work increment $\delta W_\text{fb}[\gamma](t_k)$ for the state to be stabilized.  The trajectory $\gamma$ thus gives rise to two normalized distributions 

\bb \label{eq:dist}
P_\gamma[\delta Q_\mathrm{q}] = \sum_{n=0}^{N} \delta^{\cal D} (\delta Q_\mathrm{q} - \delta Q_\mathrm{q}[\gamma](t_n) ) / Q_\mathrm{q}[\gamma],
\ee
\noindent and
\bb \label{eq:dist}
P_\gamma[\delta W_\mathrm{fb}] = \sum_{n=0}^{N} \delta^{\cal D} (\delta W_\mathrm{fb} - \delta W_\mathrm{fb}[\gamma](t_n) ) / W_\mathrm{fb}[\gamma],
\ee

\noindent which perfectly match if the feedback is optimal. Just like the energy fluctuations (Eq.\ref{eq:fluc}), the quantum heat distribution quantifies the strength of the measurement performed by the reservoir: Its typical support $|\delta Q|^\mathrm{max}$ is all the larger as the rate of decoherence $\Gamma^*$ or the Qubit's coherences $\langle \sigma_-\rangle$ are increased. This puts physical constraints on the feedback source, which must be able to provide a power ${\cal P} = |\delta Q|^\text{max}/\tau_\mathrm{fb}$ to stabilize the state. Reciprocally, a finite source power ${\cal P}^\text{max}$ leads to a cutoff in the work distribution's support $|\delta W_\text{fb}|^\text{max} = {\cal P}^\text{max} \tau_\mathrm{fb}$, eventually altering the stabilization (See Fig.\ref{fig4}). In this spirit we have also studied the dependence of the feedback's performances, as a function of the state to stabilize $\ket{\psi_\text{targ.}}$. As expected, the support of the quantum heat distribution is all the smaller as the state approaches the poles of the Bloch sphere, which are stable under the monitoring process. Therefore the feedback requires less and less power (See Fig.\ref{fig5}). Just like the response time (here taken as infinitely short), the quantum heat distribution appears as an essential tool to evaluate the quality of a feedback loop. \\

\label{conclusion}
\noindent\textbf{Conclusion}\\
It has for long been known that quantum measurement is an essential cause of stochasticity and irreversibility in a quantum system's evolution, just like the coupling to a thermal bath randomizes the evolution of classical systems. Building on this analogy, we have identified the energetic quantum fluctuations induced by measurement back-action with a genuinely quantum component of heat exchanges, and suggested a consistent framework for quantum thermodynamics, that is entirely based on standard quantum formalism. We have then used this framework to investigate energetic aspects of measurement induced irreversibility, and to provide a new thermodynamic perspective on textbook situations of quantum optics. By the way, we have shown that the concept of quantum heat is a real, physical quantity having clear operational consequences, e.g. {that} can be used to define new merit criteria to measure the quality of a feedback loop.

As it focuses on the energetic aspects of quantum measurement and decoherence, this framework brings new tools to understand and investigate the energetic cost of quantum processes, which have systematically involved a thermal bath so far. In particular, we have addressed the work cost of fighting against decoherence on a simple physical example, that can be generalized to more complex algorithms of quantum computing. The analogy between a measuring device and a Markovian bath can also be fruitfully exploited in the context of quantum engines, where the measuring apparatus can be treated as some genuinely quantum energy source \cite{zeno}. In such engines, work extraction is altered by measurement induced irreversibility, just like classical irreversibility decreases the efficiency of classical heat engines.  \\

\label{Methods}
\noindent\textbf{Methods}\\
{\footnotesize
{
\noindent\textbf{Stochastic Schr\"odinger equation in QJ and QSD.} In the QJ unraveling, the norm-preserving stochastic Schr\"odinger equation writes \cite{Wiseman-09}:

\bb
 &d\ket{\psi_\gamma(t)}& = \Bigg[-i dt H_\text{eff}^\text{(QJ)}(t) \nonumber\\
 &&+ \sum_k dN_k(t) \left( \dfrac{L_k}{\sqrt{\langle L_k^\dagger L_k \rangle}} - 1\right)\Bigg]\ket{\psi_\gamma(t)},\nonumber
\ee
where $\langle\cdot\rangle = \bra{\psi_\gamma(t)}\cdot\ket{\psi_\gamma(t)}$. We have introduced the non-hermitian Hamiltonian

$$H^\text{(QJ)}_\text{eff}(t)= H_\text{s}(t) - (i/2dt)\sum_k \Gamma_k( L_k^\dagger L_k-\langle L_k^\dagger L_k \rangle),$$
and the Poisson process $dN_k(t)$ equal to $1$ if the outcome ${\cal K}_\gamma(t)=k$ is recorded at time $t$, and $0$ otherwise.

In the QSD unraveling, the norm-preserving stochastic Schr\"odinger equation writes \cite{Brun-00}:
$$
 d\ket{\psi(t)} = \left[-i dt H_\text{eff}^\text{(QSD)}(t)+ \sum_{k} \sqrt{\Gamma_k} dw_{k}^{\cal K}(t)\left( L_k - \left\langle L_k \right\rangle\right)\right]\ket{\psi(t)}.
$$
We have introduced the non-hermitian Hamiltonian 

$$
H^\text{(QSD)}_\text{eff}(t)= H_\text{s}(t) +  i\sum_k \Gamma_k \left( \left\langle L_k^\dagger\right\rangle L_k -\tfrac{1}{2} L_k^\dagger L_k-\tfrac{1}{2} \vert\left\langle L_k^{}\right\rangle\vert^2\right).$$  

\noindent and the set of Wigner increments $\{dw_k^{\cal K}(t)\}_{1\leq k \leq {\cal N}_\text{s}^2-1}$ verifying Eq.\ref{dw_mean}-\ref{dw_var}.

\noindent\textbf{Expressions of work and heat increment operators.} The work increment operator is defined by $\delta W_\gamma = \bra{\psi_\gamma(t)} \dbar \hat{W}(t)\ket{\psi_\gamma(t)}$. For any unraveling, it corresponds to the Hamiltonian variation during $dt$: 
$$\dbar \hat{W}(t) = dH_\text{s}(t).$$
The heat increment operator is defined as $\delta Q_\gamma = \bra{\psi_\gamma(t)} \dbar \hat{Q}(t)\ket{\psi_\gamma(t)} = dU(t) - \delta W$ and depends on the unraveling. For QJ, it reads:
$$
\dbar\hat{Q}(t) =  \sum_k \bigg[ dN_k(t) \dfrac{L_k^\dagger \Delta_s(t) L_k}{\left\langle L_k^\dagger L_k\right\rangle} - \dfrac{\Gamma_kdt}{2}\left(L_k^\dagger L_k\Delta_s(t) + \text{H.c.}\right)\bigg],
$$
where $\text{H.c.}$ stands for hermitic conjugate and $\Delta_\text{s}(t) = H_\text{s}(t) - U_\gamma(t)$. For QSD it takes the form:

\bb
\dbar\hat{Q}(t)&=&  \sum_k \bigg[ \sqrt{\Gamma_k} dw_k^{\cal K}(t) (L_k^\dagger \Delta_\text{s}(t)  +  \Delta_\text{s}(t) L_k)\nonumber\\
&&+ \Gamma_k dt\left(L_k^\dagger H_\text{s}(t) L_k- \tfrac{1}{2}\{L_k^\dagger L_k, H_\text{s}(t)\}\right)\bigg].\nonumber
\ee
}
\\

\noindent\textbf{Spontaneous emission of a Qubit}\\
{
\noindent\emph{Heat}. As the system Hamiltonian is time-independent, the work increment is null at any time: $\delta W = 0$. Using the heat operator for QJ unraveling, we find that during a jump (measurement outcome $m_{{\cal K}(t)} = m_1$), the heat increment is $\delta Q = - \hbar \omega_0 P_e(t)$, with $P_e(t) = \bra{\psi_\gamma(t)}\sigma^\dagger\sigma\ket{\psi_\gamma(t)}$ the population of the excited atomic level. Using the decomposition presented under Eq.\ref{eq:QJ}, we can write $\delta Q/\hbar  = -\omega_0+ \omega_0(1-P_e(t))$, where the first term is the classical heat $Q_\text{cl}(1)$ carried by the emitted photon, and the second quantifies the energy fluctuation (quantum heat) due to the Qubit's measurement in $\ket{e}$. When no jump occurs ($m_{{\cal K}(t)} = m_0$), we find $\delta Q = - \hbar \omega_0dt P_e(t)(1-P_e(t)) = \delta Q_q$. \\

\noindent\emph{Entropy production}.
The direct trajectory starts in $\ket{+_x}$ with probability $p_\text{d}(+_x) = 1$. The conditional probability $P_\text{d}[\gamma\vert +_x]$ is computed using Eq.\ref{eq:ptraj}. For the no-jump trajectory $\gamma_\text{nj}$, it writes:
\bb
p_\text{nj}(t) &=& \left\vert\bra{\psi_\text{nj}(t)} \prod_{n=1}^N (1-idt\omega_0\sigma_+\sigma_- -\Gamma dt \sigma_+\sigma_- /2) \ket{+_x}\right\vert^2\\
					&=& \left\vert\bra{\psi_\text{nj}(t)} \exp(-(i \omega_0+\Gamma)t \sigma_+\sigma_-/2) \ket{+_x}\right\vert^2,
\ee
leading to $p_\text{nj}(t) = (1+e^{-\Gamma t})/2$. The time-reversed no jump trajectory starts in $\ket{\psi_\text{nj}(t_N)}$ with probability $p_\text{r}(\psi_N) = P_\text{d}[\gamma_\text{nj}\vert +_x]$ and ends in $\ket{+_x}$, yielding eventually:
 
\bb
P_\text{r}[\gamma_\text{nj}^\text{r}\vert \psi_f] &=& \left\vert\bra{+_x} \exp((i \omega_0-\Gamma)t \sigma^\dagger\sigma/2) \ket{\psi_\text{nj}(t_N)}\right\vert^2 = p_\text{nj}(t).
\ee
Therefore $\Delta_\text{i}^c s[\gamma_\text{nj}] = 0$, and $\Delta_\text{i}^\text{b} s[\gamma_\text{nj}] = \Delta_\text{i} s[\gamma_\text{nj}] = \log(2/(1+e^{-\Gamma t}))$.
For a trajectory $\gamma_j$ featuring a jump at time $t_j$, with $t_j \leq t$, the probability of the trajectory reads:
\bb
&&P_\text{d}[\gamma_j\vert +_x] = \Gamma dt \nonumber\\
&&\times\bigg\vert\bra{g} e^{-(i\omega+\Gamma)(t-t_j)\sigma_+\sigma_-/2}\sigma e^{-(i\omega+\Gamma)t_j)\sigma_+\sigma_-/2} \ket{+_x}\bigg\vert^2\quad.
\ee
leading to $p_j(t) = \Gamma dt e^{-\Gamma t_j}/2$. The time-reversed trajectory $\gamma_j^\text{r}$ starts in state $\ket{g}$ with probability $p_\text{j}(t) = 1-p_\text{nj}(t)$ and involves the time-reversed operator of $\sigma_-$, which is given by Eq.\ref{Mrbath}, with $Q_\text{cl}({\cal K}) = - \hbar \omega_0$. In the zero temperature limit considered in this example, $M_{\cal K}^r = 0$: Therefore $P_\text{r}[\gamma_j^\text{r}\vert +_x] = 0$, and $\Delta_\text{i}^\text{c} s[\gamma_j]$ diverges. The boundary term reads: $\Delta_\text{i}^\text{b}s[\gamma_\text{j}] = \log(2/(1-e^{-\Gamma t}))$.\\

\noindent\textbf{Weak monitoring of a Qubit}
The increment of internal energy can be computed using $d(\bra{\psi_\gamma(t)}H\ket{\psi_\gamma(t)}) = d\bra{\psi_\gamma(t)}H\ket{\psi_\gamma(t)}+\bra{\psi_\gamma(t)}H d\ket{\psi_\gamma(t)} + d\bra{\psi_\gamma(t)} H d\ket{\psi_\gamma(t)}$, where the last term has to be computed up to order $dt$ using Ito's rules, i.e. $dw_\gamma(t)^2 = dt$, $dw_\gamma(t)dt = 0$. The Hamiltonian of the Qubit writes $H=\hbar \omega_0\sigma_z/2$, such that

\bb
dU_\gamma &=&\hbar  \omega_0\sqrt{\Gamma^*}dw_\gamma(t)\left(1-(\bra{\psi_\gamma(t)}\sigma_z\ket{\psi_\gamma(t)})^2\right)\nonumber\\
  &=& 4 \hbar \omega_0\sqrt{\Gamma^*}dw_\gamma(t)\vert\bra{\psi_\gamma(t)}\sigma_-\ket{\psi_\gamma(t)}\vert^2,
\ee
where we have used the identity $1-\langle\sigma_z\rangle^2 = 4\vert\langle\sigma_-\rangle\vert^2$.\\
}

\begin{acknowledgements}\noindent \textbf{Acknowledgments -- } It is a pleasure to thank S. T. Bramwell, L. Bretheau, P. Campagne-Ibarcq, P. Degiovanni, I. Fr\'erot, P. C.W. Holdsworth, B. Huard, D. Lacoste, A. Mitra, J.-M. Raimond, N. Roch, B. Roussel and T. Ziman for discussions and comments. M.C. thanks the Centre de physique th\'eorique Grenoble-Alpes and the theory group at Institut Laue-Langevin for hospitality during completion of this work. Part of this work was supported by the COST Action MP1209 ``Thermodynamics in the quantum regime" and ANR grant No. ANR-13-JCJC-INCAL. \\

\noindent {\it We dedicate this manuscript to Maxime Clusel who initiated this work.}\\

\noindent\textbf{Correspondence --} Correspondence and requests for materials should be addressed to Alexia Auff\`eves (alexia.auffeves@neel.cnrs.fr).
\end{acknowledgements}

  }

\end{document}

\title{The role of quantum measurement in stochastic thermodynamics\\
Supplementary material}
\author{Cyril Elouard}
\author{David A. Herrera-Mart\'i}
\affiliation{Institut N\'eel, UPR2940 CNRS and Universit\'e Grenoble Alpes, avenue des Martyrs, 38042 Grenoble, France.}
\author{Maxime Clusel}
\affiliation{Laboratoire Charles Coulomb, UMR5221 CNRS and Universit\'e de Montpellier, place E. Bataillon, 34095 Montpellier, France.}
\author{Alexia Auff\`eves}
\affiliation{Institut N\'eel, UPR2940 CNRS and Universit\'e Grenoble Alpes, avenue des Martyrs, 38042 Grenoble, France.}
\date{\today}
\pacs{
05.30.-d, 05.40.-a, 05.70.Ln, 05.30.-d, 42.50.Lc
}
\maketitle

\section*{A microscopic interpretation of the quantum heat}

Here we provide a simple interpretation of the quantum heat contribution in the context of cavity quantum electrodynamics. We consider a Qubit of transition frequency $\omega_0$ prepared in the coherent superposition $\ket{+_x}=(\ket{e}+\ket{g})/\sqrt{2}$, and then measured in its energy eigenbasis. Preparation and measurement fully define the thermodynamic transformation under study. For this elementary two-points quantum trajectory, the internal energy change equals $\Delta U = Q_\text{q}= \pm  \hbar \omega_0/2$. 

It is possible to account for this energy change by introducing the classical source used to prepare the Qubit in a coherent superposition {\it before} the measurement. It is modeled by a large coherent field $\ket{\alpha}$ injected in the mode of a high quality factor cavity, resonantly coupled to the Qubit. Starting from the state $\ket{g,\alpha}$, the Qubit-field system evolves by coherently exchanging a quantum of energy (Rabi oscillation). After a time $\tau$, the Qubit and the field share an entangled state ${\cal{N}}(\tau)(\ket{e,\alpha_e(\tau)}+\ket{g,\alpha_g(\tau)})$, where ${\cal N}(\tau)$ is a normalisation constant. $\ket{\alpha_e}$ (resp. $\ket{\alpha_g}$) is the field's state correlated with the Qubit's excited (resp. ground) state, checking $\langle \alpha_e(\tau_{1/2}) |N| \alpha_e(\tau_{1/2}) \rangle = \langle \alpha_g(\tau_{1/2}) |N| \alpha_g(\tau_{1/2}) \rangle -1$, where $N$ is the photon number operator. In the classical limit where $|\alpha|\gg 1$, we have $\alpha_e(\tau) \sim \alpha_g(\tau)$ such that after tracing over the field, the Qubit remains in a pure state. This quantum-classical boundary has been theoretically studied by Gea-Banacloche \cite{Gea-90}, and experimentally checked by Auff\`eves \textit{et al} \cite{Auffeves-03}. 

An equally weighted coherent superposition of $\ket{e}$ and $\ket{g}$ is prepared if the Qubit-field interaction is frozen at time $\tau_{1/2}$, such that $P_e(\tau_{1/2}) = P_g(\tau_{1/2})$. $P_e$ (resp. $P_g$) are the excited (resp. ground) state populations. Before the measurement, the quantum of energy is equally delocalized between the Qubit and the source, corresponding to the Qubit's energy change $\Delta U=\hbar \omega_0/2$. After the measurement, the Qubit is projected on the state $\ket{e}$ (resp. $\ket{g}$). Simultaneously, the field is projected on the state $\ket{\alpha_e(\tau_{1/2})}$ (resp. $\ket{\alpha_g(\tau_{1/2})}$). In the first case, the quantum of energy is localized in the Qubit, and in the source in the second case. The quantum heat accounts for this energy relocalization. 

As mentioned in the main text, each physical situation can give rise to such analysis involving the remote sources used to prepare the superposition. However, these sources are most often not accessible; Moreover, in the classical limit of large numbers of photons in the field the degree of atom-field entanglement is negligible as the atomic purity approaches $1$. This motivates the strategy proposed in the paper, i.e. not searching for further justifications of the quantum heat, and take it as a natural byproduct of the standard measurement postulate.

\section*{Quantum Jarzynski's equality}

Just like in the classical situation, the entropy produced by some quantum measurement and its energetic imprint, e.g. the quantum heat, can be quantitatively related in the quantum Jarzynski's protocol. It involves a system initially thermalized in a bath of inverse temperature $\beta$, then decoupled from the bath. Its energy is measured with outcome $U(t_0)$, preparing the corresponding energy eigenstate. It is then driven by a Hamiltonian $H_\mathrm{s}(t)$, and its energy measured again with outcome $U(t_N)$, giving rise to a stochastic change of internal energy $\Delta U[\gamma]=U(t_N)-U(t_0)$. It is well known that 
\bb
\langle \exp(-\beta(\Delta U[\gamma]-\Delta F))\rangle_\gamma = 1\label{res:JE}, 
\ee
where $\langle\cdot\rangle_\gamma$ stands for the average over all trajectories and $\Delta F$ is the change in the system's free energy. With current definitions, the internal energy change is solely attributed to work exchanges: $\Delta U[\gamma]=W[\gamma]$. This gives rise to the quantum version of Jarzynski's equality, whose formal similarity with its classical counterpart has long been known. In particular, the entropy produced by the trajectory $\gamma$ has the same expression as in the classical case $\Delta_\mathrm{i}s[\gamma] = \beta(W[\gamma]-\Delta F) = \beta Q_\mathrm{irr}[\gamma]$, where $Q_\mathrm{irr}[\gamma]$ is classically defined as the heat irreversibly dissipated in the bath.  However, the physical situations addressed by each protocol are clearly different: While the classical protocol allows measuring irreversibility due to the departure from thermal equilibrium, the quantum protocol can involve purely quantum irreversibility, due to coherences' erasure induced by the final energy measurement \cite{Manzano-15,Plastina-14}. Such quantum irreversibility takes place if the system is driven in a non-adiabatic way, such that its state before the final measurement is not an energy eigenstate.  Within the current framework, entropic contributions of measurement-induced irreversibility have remained hidden, while they clearly appear with our definitions: The change of internal energy now reads $\Delta U[\gamma] = W[\gamma] + Q_\mathrm{q}[\gamma]$, where $W[\gamma] =U(t_N^{-}) - U(t_0)$ is the work deterministically exchanged with the drive, and $Q_\mathrm{q}[\gamma] = U(t_N)-U(t_N^{-})$ the stochastic heat exchange due to measurement back-action. Finally, the entropy production now splits into two components:
\bb
\Delta_\text{i}s[\gamma]&=& \beta(\Delta U[\gamma]-\Delta F)\nonumber\\
&=& \beta(W[\gamma]-\Delta F)+\beta Q_\mathrm{q}[\gamma].\label{res:irr}
\ee
The term $\beta Q_\mathrm{q}[\gamma]$ is a purely quantum contribution, that quantitatively relates the energetic and entropic signatures of quantum irreversibility. This term vanishes in the classical situation where the energy eigenbasis is fixed, or changes slowly (adiabatically). Note that if both quantities appear in similar situations of non-adiabatic driving, the quantum heat is not equivalent to the internal friction heat $Q_\text{fric}$ \cite{Plastina-14}, which is usually interpreted as the heat that would be dissipated in a fictitious heat bath and is strictly proportional to entropy production. Here the quantum heat can be negative, and corresponds to a real system's energy change that takes place during the measurement induced quantum jump.

We now show that the expression \ref{res:irr} for entropy production can be extended to the case of quantum open systems.

\section*{Jarzynski's equality for a quantum open system.}

Here we consider a driven quantum system coupled to a thermal bath of inverse temperature $\beta$. This situation is routinely described by a Lindbladt master equation, that is unraveled within a QJ picture. Each jump operator induces a single transition between two states of a given basis ${\cal B}=\{\ket{i}\}$, described by Eq.(16) of main text.  We compute the probability of a direct trajectory $\gamma$ and the corresponding reverse trajectory $\gamma_\text{r}$, applying formul\ae  17 and 18 of main text. $\ket{\psi_0}$ (resp. $\ket{\psi_N}$) is the initial (resp. final) energy eigenstate in which the system is just after the initial (resp. final) energy measurement. The boundary term in the entropy production is identical to the closed system case, verifying $\Delta_\text{i}^\text{b}s[\gamma]=-\log(p_\text{d}(\psi_N)/p_\text{d}(\psi_0)) = e^{-\beta (\Delta U[\gamma]-\Delta F)}$. On the other hand, the contribution of the conditional probabilities is now non-trivial. The equilibrium state $\pi_\text{s}$ of the Lindbladian is a thermal state verifying
\bb
\pi_\text{s}=\sum_{i=1}^{{\cal N}_s} \pi_\text{s}(i) \ket{i}\bra{i},
\ee 
with $\pi_\text{s}(i) = \exp(-\beta (U_i(t_n)-F(t_n)))$. $U_i(t_n)$ is the energy of state $\ket{i}$ and $F(t_n)$ the equilibrium free energy. Therefore, the reversed measurement operators $M_{\cal K}^\text{r}$ verify Eq.20 of main text with $Q_\text{cl}({\cal K}) = U_{j({\cal K})}-U_{i({\cal K})}$.  
Finally, the conditional probability of the reversed trajectory reads:
\begin{widetext}
\bb
P_\mathrm{r}[\gamma_\mathrm{r}|\psi_N] &=& |\bra{\psi_0} \left(\prod_{n=1}^N  \exp\left[\beta Q_\text{cl}({\cal K}_\gamma(t_{N+1-n}))\right] M^\dagger_{{\cal K}_\gamma(t_{N+1-n})}\right)\ket{\psi_N} |^2 \nonumber\\
&=& \exp\left[\beta \sum_{n=1}^N Q_\text{cl}({\cal K}_\gamma(t_{N+1-n}))\right] |\bra{\psi_0} \left(\prod_{n=1}^N  M^\dagger_{{\cal K}_\gamma(t_{N+1-n})}\right)\ket{\psi_N} |^2\nonumber\\
&=& \exp\left[\beta Q_\text{cl}(\gamma)\right]|\bra{\psi_N} \left(\prod_{n=1}^N  M_{{\cal K}_\gamma(t_{n})}\right)\ket{\psi_0} |^2.
\ee
\end{widetext}
We have used the equalities $\sum_{n=1}^N Q_\text{cl}({\cal K}_\gamma(t_n)) = Q_\text{cl}[\gamma]$, and $\left(\prod_{n=1}^N  M^\dagger_{{\cal K}_\gamma(t_{N+1-n})}\right)^\dagger = \prod_{n=1}^N  M_{{\cal K}_\gamma(t_{n})}$. It is then straigthforward to derive $P_\mathrm{r}[\gamma_\mathrm{r}|\psi_N] = e^{\beta Q_\text{cl}[\gamma]}P_\mathrm{d}[\gamma|\psi_0]$. We eventually get:
\bb
\Delta_i s[\gamma]= \beta\left(\Delta U - Q_\text{cl}[\gamma]-\Delta F\right).
\ee
Using the first law at the single trajectory level:
\bb
\Delta U = W[\gamma]+Q_\text{q}[\gamma]+Q_\text{cl}[\gamma],
\ee
we finally find
\bb
\Delta_i s[\gamma] = \beta\left(W[\gamma]-\Delta F\right)+\beta Q_\text{q}[\gamma],
\ee
which extends the validity of Eq.\ref{res:irr} (See previous section). Again, the second term involving the quantum heat contribution quantifies the irreversibility induced by the quantum measurements performed on the system. Such measurements involve the continuous monitoring of the environment and the final energy measurement performed by the operator.

In particular, this formalism captures genuinely quantum situations for which the time-dependent Hamiltonian $H_\mathrm{s}(t_n)$ does not commute with the stationary state defined by the Lindbladian $\pi_\text{s}(t_n)$ at the same time \cite{Elouard-16, massi}. Here $\beta Q_\text{q}$ is non-zero because after each jump, the system is projected on one of the states of ${\cal B}$, i.e. an eigenstate of $\pi_s$, which is not an eigenstate of the Hamiltonian $H_\text{s}$: It therefore develops coherences during the deterministic evolution, before a new jump takes place. Again, this expression for entropy production is fully compatible with current derivations of FTs: The main difference of our approach is that the heat increment now has a classical and a quantum component, and solely the classical component is involved in the expression of the time-reversed operators.\\